# Molecular Modelling Combined with Advanced Chemistry for the Rational Design of Efficient Graphene Dispersing Agents

Konstantinia D. Papadimitriou,[a,b] Emmanuel N. Skountzos,[a,b] Sandra S. Gkermpoura,[a,b] Ioannis Polyzos,[a] Vlasis G. Mavrantzas,[a,b,c] Costas Galiotis[a,b] and Constantinos Tsitsilianis[a,b,*]

[a]Foundation of Research and Technology Hellas, Institute of Chemical Engineering Sciences (FORTH/ICE-HT), Stadiou Str., P.O. Box 1414, GR 26504, Rio-Patras, Greece

[b]Department of Chemical Engineering, University of Patras, GR 26504, Patras, Greece

[c]Particle Technology Laboratory, Department of Mechanical and Process Engineering, ETH-Z, CH-8093 Zürich, Switzerland

**ABSTRACT:** Pyrene-functional PMMAs were prepared via ATRP-controlled polymerization and click reaction, as efficient dispersing agents for the exfoliation of few-layered graphene sheets (GS) in easily processable low boiling point chloroform. In parallel, detailed atomistic simulations showed fine dispersion of the GS/polymer hybrids in good agreement with the experiment. Moreover, the molecular dynamics simulations revealed interesting conformations (bridges, loops, dangling ends, free chains) of GS/polymer hybrids and allowed us to monitor their time evolution both in solution and in the polymer nanocomposite where the solvent molecules were replaced with PMMA chains. Microscopic information about these structures is very important for optimizing mechanical performance. It seems that the combination of atomistic simulation with advanced chemistry constitutes a powerful tool for the design of effective graphene dispersing agents that could be used for the production of graphene-based nanocomposites with tailor-made mechanical properties.

Graphene has attracted a lot of attention recently because of its extraordinary combination of properties that render it an extremely promising material for several applications.[1-3] Among different approaches to produce graphene, liquid-phase exfoliation methods are widely considered to be a promising route for large-scale and defect-free graphene production due to their scalability and versatility for a wide range of applications including polymer composites.[4-7] One of the problems that must be tackled in this context is the tendency of graphene sheets (GS) to agglomerate into multilayer graphitic structures due to the presence of van der Waals forces and π–π interactions. Hence, it is of particular importance to prevent the graphene sheets from aggregating because most of its unique properties emanate from its two-dimensional character rather than that of multilayer graphite.

To overcome this problem, functionalization of graphene, through covalent or non-covalent modifications, has been found to improve its dispersion in solvents and its processability, thus, broadening its applications.[8] Among these methods, non-covalent functionalization of graphene through π–π interactions as the binding forces between graphene and stabilizers is the most effective and non-destructive method, which enables the modification of material properties without altering the chemical structure of graphene. To this direction, pyrene derivatives and pyrene-functional polymers were prepared and evaluated either as graphene exfoliation agents or for the fabrication of nanostructure composites by self-assembly.[9-13] The adsorption of pyrene derivatives on graphene through robust π-π stacking interactions led to highly uniform and stable dispersions of graphene without inflicting any damage to the graphitic surface.[14-16] The main advantage of using pyrene-polymers as modifiers for the functionalization of graphene is their good compatibility with different polymer hosts, which renders graphene a versatile nanofiller in composites.

Dispersing graphene in solution constitutes the first step towards nanocomposite fabrication.[17] However, efficient dispersion in various liquids does not ensure that the dispersibility is retained within the polymer matrix after solvent removal. In this work, we have embarked on new chemical strategies and synthetic routes to avoid re-agglomeration of isolated graphene sheets, an approach also confirmed and supported by explicit atomistic simulations. In particular, the present communication demonstrates a method for the liquid-phase exfoliation of graphene in an easily processable low-boiling point solvent ($CHCl_3$) and discusses the structure of the resulting graphene/PMMA nanocomposites (i.e., after the removal of the solvent). For this purpose, non-covalent functionalization of the basal plane of graphene was attempted by designing welldefined pyrene end-capped PMMA (Py-PMMA-Py) as the dispersing agent. High concentrated stable graphene dispersions in chloroform were obtained by implementing various exfoliation procedures and strategies. In parallel to the synthetic work, atomistic-level simulations were carried out in order to better understand the liquid exfoliation process, to reveal the GS/polymer hybrid conformations, and to assess whether the Py-PMMA-Py compound can prevent the agglomeration of graphene in the PMMA matrix, as occurs for non-functionalized PMMA.[18] The simulations showed not only the capability of the designed polymer to prevent GS agglomeration within the PMMA matrix but also, and more importantly, revealed various conformations (e.g., GS bridged by PMMA) which could have a strong effect in enhancing the mechanical properties of the nanocomposite.

For the non-covalent functionalization of graphene, a pyrene end-capped PMMA was prepared following the multistep procedure demonstrated in Scheme 1 (see experimental details in Supporting Information). In particular, a suitable py-functional initiator (1) was first synthesized and used to prepare α-functional Py-PMMA, the end of which was subsequently modified to azide function, yielding an α-ω heterotelechelic Py-PMMA-$N_3$ (2). The synthesis of Py-PMMA-Py, was accomplished via the azide-alkyne click reaction. The Py-PMMA-$N_3$ reacted with a prior synthesized alkyne-functional pyrene derivative (3) yielding the Py-PMMA-Py (4) final product through 1,2,3 triazole.

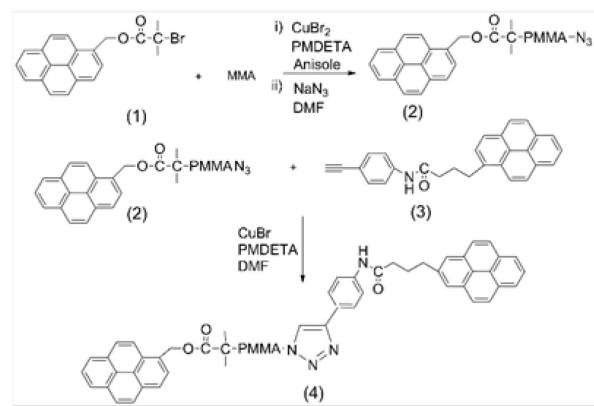

**Scheme 1**. Synthesis of pyrene-functional α,ω-PMMA using the ARGET-ATRP polymerization and click chemistry.

A well-defined pyrene-functional PMMA of low polydispersity was thus obtained as confirmed by GPC and $^1$H-NMR (see SI). As seen in Table 1, there is an excellent agreement between the $^1$H-NMR and GPC in $M_n$ data.

To estimate the pyrene end-functionalization, UV–vis measurements were also performed (Fig. S1) indicating quantitative pyrene ω-functionalization of Py-PMMA. It should be noted that the synthetic strategy adopted herein ensures quantitative pyrene end-capping of PMMA.

**Table 1. Characteristics of the pyrene-functional PMMA**

| Polymer | $DP_n$ [a] | $M_n$ [a] | $M_n$ [b] | $M_w$ [b] | PDI [b] |
|---|---|---|---|---|---|
| Py-PMMA | 50 | 5410 | 5074 | 5774 | 1.13 |
| Py-PMMA-Py | 50 | 5730 | 5389 | 6116 | 1.13 |

Determined by: [a] $^1$H NMR, [b] GPC

As already mentioned earlier, the capability of Py-PMMA-Py to prevent GS agglomeration in the presence of the solvent was directly confirmed by detailed, all-atom molecular dynamics (MD) simulations with a model system (called System 1 here) containing 3061 chloroform molecules, 40 Py-PMMA-Py chains ($DP_n$ 30) and 6 hydrogen-terminated GS with lateral dimensions 39 Å × 39 Å, in a simulation cell subject to periodic boundary conditions along all 3 space directions. The simulation was executed with the LAMMPS software[19] at $T$ = 300 K and $P$ = 1 atm and all technical details can be found in the Supporting Information (S.I.). No GS agglomeration was observed after several hundreds of nanoseconds of MD simulation (see Figure 1a). A detailed conformational analysis of the simulation trajectory revealed that pyrene end-functional groups tend to adsorb strongly on GS, leading to the following characteristic structures (see Figure 1b-e): a) dangling ends (formed by py-PMMA-py chains that adsorb on a graphene sheet by only one of their end pyrene groups, thus, leaving the other end free, see Figure 1b); b) loops (formed py-PMMA-py chains adsorbed by both of their end pyrene groups on the same face of a GS, see Figure 1d); c) extended loops (formed by py-PMMA-py chains adsorbed on the two different faces of the same GS, see Figure 1c); and d) bridges (formed py-PMMA-py chains simultaneously adsorbed on two different GS, see Figure 1e). These adsorbed structures prevent GS self-assembly in the form of π-π stacks. Of course, we also observed py-PMMA-py chains with both of their ends free (i.e., non-adsorbed chains). In Figure 1a-e, carbon atoms of graphene sheets and pyrene molecules have been colored in yellow and black, respectively, while carbon atoms belonging to the backbone of completely free, dangling end, loop and bridge conformations have been colored in white, green, blue and red, respectively. The rest of atoms (hydrogens and atoms belonging to the side groups of Py-PMMA-Py) and the chloroform molecules are not shown for clarity. That GS do not exhibit any tendency to self-assemble was further confirmed by examining the evolution in time of the distance between the centers-of-mass of all GS pairs present in the simulation cell (see Figure S5). In all cases, these were calculated to be significantly larger than the characteristic equilibrium distance of successive GS in a graphite flake (= 3.4 Å). Our results agree with previous MD studies by Xu and Yang,[20] and the single molecule force spectroscopy study of Zhang et al.,[21] which provided strong evidence for a very favourable interaction of mono-functional pyrene–polyethylene glycol (Py–PEG) with GS. The time evolution of the % percentage of the different types of adsorbed configuration is given in Figure 1f. As observed, during the first ~10ns of the simulation the number of free chains decreases dramatically while the corresponding number of dangling ends increases. At ~20ns, the number of dangling ends reaches a peak and then starts to decrease. In contrast, the number of loops and bridges increases continuously towards their asymptotic values at very long times. The most probable conformation is that of bridges amounting to about 50% of all (adsorbed and non-adsorbed) Py-PMMA-Py structures. We also observe a good percentage of loops and dangling ends, and only a very small percentage (~5%) of free chains. Among all these conformations, bridges of high density are the most important ones since they are expected to have a pronounced effect on the elastic modulus in PMMA nanocomposite or ion gels after replacing $CHCl_3$ with an ionic liquid.[22] In this respect, atomistic simulations seem to be an invaluable tool capable of determining the optimal GS/polymer ratio, that is, the one that maximizes the number of bridges for the final nanocomposite to exhibit superior mechanical performance.

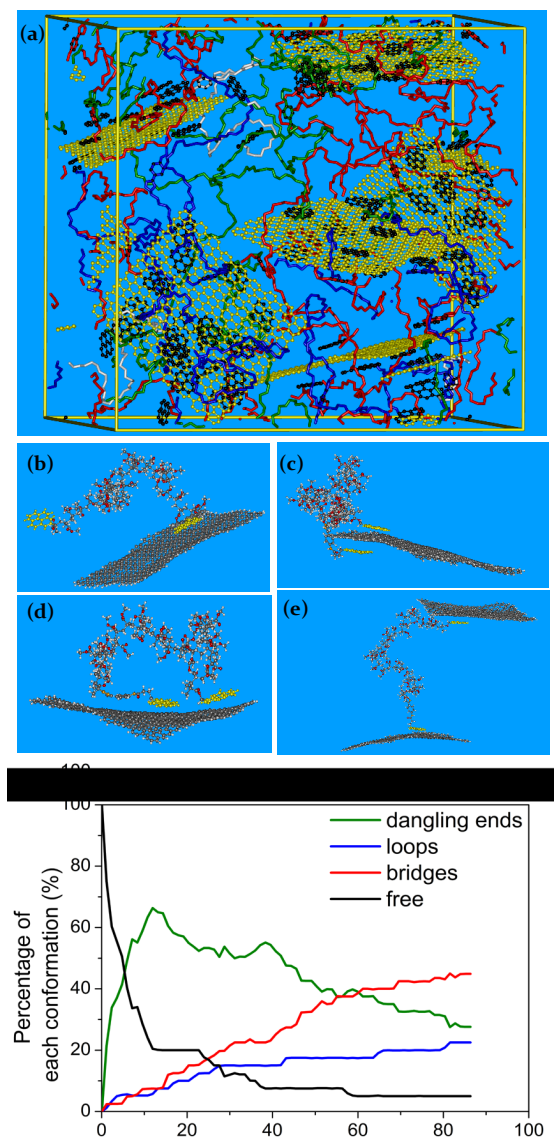

**Figure 1.** (a) Typical conformations of adsorbed and free Py-PMMA-Py chains on GS (in yellow) in the simulation cell after several nanoseconds of simulation time. Different conformations are shown in different colors: loops in blue, bridges in red, dangling ends in green, and free chains in white. b-e) Examples of typical Py-PMMA-Py conformations adsorbed on GS. f) Time evolution of the relative population of adsorbed (dangling ends, loops, bridges) and non-adsorbed (free) Py-PMMA-Py conformations in the course of the *NPT* MD simulation.

The liquid exfoliation of graphite in chloroform in the presence of Py-PMMA-Py as stabilizer was subsequently explored. The selection of chloroform as a solvent is due to its low boiling point, which ensures easy removal, hence, facilitating the fabrication of polymer composites. To this end, a systematic study was undertaken to define the required preparation parameters, such as the exfoliation method, the centrifugation rate (ω, rpm), and the Py-PMMA-Py/feed graphite (P/GF) mass ratio. The methods associated with the process and the characterization of graphene products are described in detail in the Supporting Information. Keeping in consideration that high-shear mixing has been already reported as an effective method for graphite exfoliation[2,23] (even more efficient than tip sonication), we herein demonstrate an alternative method of exfoliation by employing instantaneously shear mixing and tip sonication (Figure S9a).

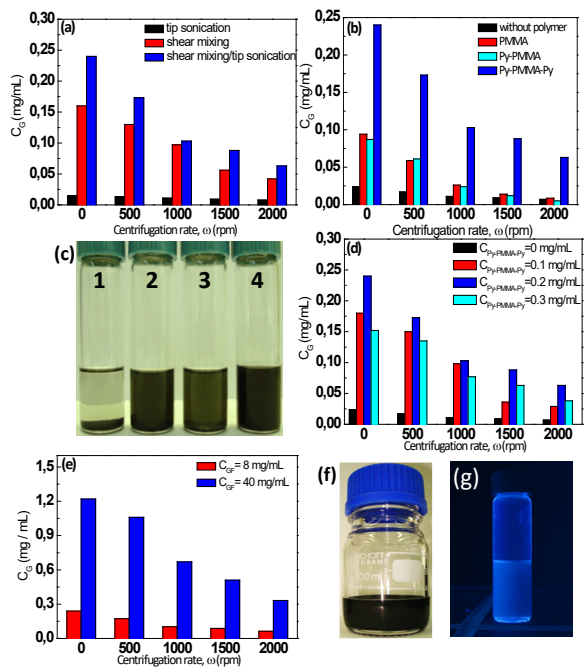

**Figure 2.** (a) Graphene concentration, $C_G$, in $CHCl_3$ from different methods of graphite exfoliation at various centrifugation rates. (b) $C_G$ without and in the presence of different polymers. (c) Dispersions of graphite in chloroform without polymer (1) and in the presence of PMMA (2), Py-PMMA (3) and Py-PMMA-Py (4) (in all cases: feed concentration of polymers was 0.2 mg/mL and for graphite, $C_{GF}$, 8 mg/mL). (d) $C_G$ as a function of feed concentration of Py-PMMA-Py at different centrifugation rates. (e) $C_G$ as a function of $C_{GF}$ at different centrifugation rates (P/GF=0.025). (f) Digital image of G/Py-PMMA-Py dispersion (solvent volume 30 ml) (g) G/Py-PMMA-Py dispersion emitting bright blue color.

As seen in Figure 2a, much higher graphene concentrations ($C_G$) were achieved with this method in comparison to other preparation routes. To evaluate the importance of the telechelic Py-PMMA-Py topology to improve graphene exfoliation in $CHCl_3$, a comparative study was carried out by using as dispersing agent a non-functionalized PMMA, the py-PMMA precursor and the Py-PMMA-Py with similar molecular weights (Figure 2b).

The $C_G$ corresponding to Py-PMMA-Py was remarkably higher (more than 166%) than the other cases for all centrifugation rates (ω, rpm) examined. Additionally, the dispersion in the presence of Py-PMMA-Py exhibited high stability since no agglomeration was observed after several weeks, as shown in Figure 2c. It is important to stress that under specific experimental conditions, and particularly, in the absence of stabilizers, $CHCl_3$ appeared to be an ineffective solvent for graphite exfoliation, leading to very low graphene concentrations due to its unfavorable surface tension.[24] Therefore, the designed telechelic Py-PMMA-Py topology proved quite efficient in producing stable graphene dispersions in chloroform. We should also note that, the G/Py-PMMA-Py hybrids emit bright blue color under UV light (365 nm; Figure 2g), exhibiting strong fluorescence properties due to the free Py-function of the dangling ends (SI).

To investigate the critical concentration of Py-PMMA-Py needed to achieve the optimum concentration of graphene, graphene dispersions were prepared with different initial Py-PMMA-Py concentrations (Figure 2d). It was observed that the concentration of graphene initially increases with the addition of polymer and then decreases. The optimal mass ratio of P/GF was determined to be 0.025 which is favorably low. Very recently, Yan and coworkers[25] reported the use of a telechelic pyrene-capped polystyrene (PyPS) for exfoliation of graphite. Under quite similar experimental conditions (exfoliation time of 1h in the presence of polymer, centrifugation rate at 1000 rpm), the $C_G$ was determined to find 0.067 mg/mL by taking PyPS/feed graphite mass ratio to be 0.2. Compared to our work, the $C_G$ was twice lower, and PyPS/feed graphite mass ratio was found to be much higher. In our attempt to scale-up production of graphene, we increased the feed graphite concentration (40 mg/mL) by keeping constant the optimum P/GF mass ratio at 0.025 (Figure 2e,f). The $C_G$ increased drastically in the range of 0.25-1.2 mg/mL, as the centrifugation rate decreases. It should be mentioned that, for the production of quite homogeneous graphene–polymer composites, the solvent processing method is compared favorably to other methods.[7,26] Nevertheless, the difficulty in removing the residual solvent[27] has a detrimental effect on the properties of the composite materials. Therefore, the fact that we obtained high graphene concentrations using a solvent of high volatility has a significant potential toward automated production of graphene-based composites.

In order to evaluate the exfoliation efficiency of graphene dispersions, transmission electron microscopy (TEM) and Raman spectroscopy were employed. Figure 3a displays representative TEM images that reveal a large quantity of few layer-graphene sheets. The majority of the flakes have lateral sizes of the order of 1 μm. Some larger flakes having lateral dimensions above 1 μm were also observed. The folding structures of graphene sheets indicate the nature of graphene to be thermodynamically stable.[28] Furthermore, Raman spectroscopy supports the TEM analysis (Figure 3b). The typical G and 2D peaks are clearly visible as well as the D and D' peaks. The G peak corresponds to the well-known $E_{2g}$ phonon and the 2D (or G') peak is the second order of the D peak. D and D' peaks correspond to defect activated phonons. It must be mentioned that the intense, in some spectra, D line is attributed mainly to edge effects since the size of the flakes is comparable with the laser spot size (~1μm). Raman Spectroscopy has also been used to estimate graphene thickness through the measurement of number of layers (up to five).[29,30]

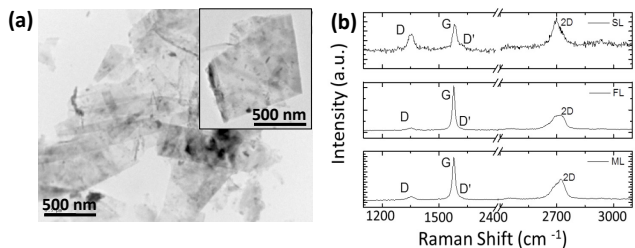
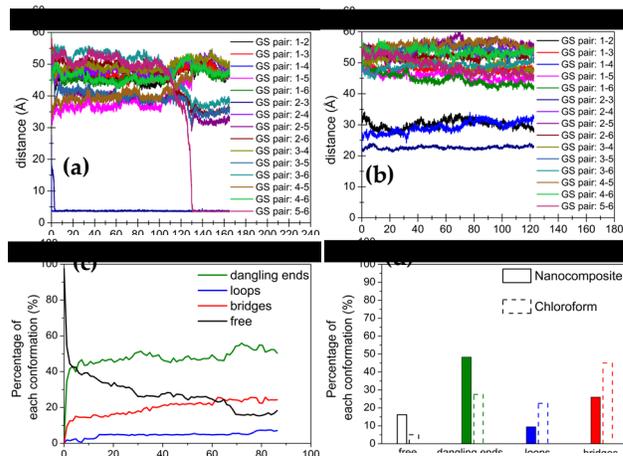

**Figure 3.** (a) TEM images of graphene/Py-PMMA-Py dispersion after centrifugation at 500 rpm (Scale bar: 500 nm; Inset: folded few-layer graphene nanosheets). (b) Representative Raman spectra of single-layer graphene (SLG), few-layer graphene (FLG) and multi-layer graphene (MLG).

Hence based on the characteristics of the 2D peak (position, shape, width, etc.) and the intensity ratio $I_{2D}/I_G$ we confirmed the presence of single layer graphene flakes at 1-2% percentage. The rest of graphene sheets were either few-layer graphenes (2-5 layers, ~50%) or multi-layer ones (6-10 layers, 25%) and, finally, nanographites (~23%). Not achieving high monolayer content is not necessarily a disadvantage since few layer GS can also be effective from the mechanical property point of view since, as argued elsewhere,[31] you can build a higher proportion of graphene into the composite and thus achieve higher values of composite modulus and strength.

To answer the question whether or not the fine dispersion of GS/polymer hybrids achieved in solution is preserved in the PMMA nanocomposites, the previous MD simulations were extended by replacing the solvent molecules with PMMA chains. Thus, two additional MD simulations were performed: In the one, the simulation cell (called System 2) contained 100 non-functionalized PMMA chains each 30-monomers long and 6 hydrogen-terminated GS (39 Å × 39 Å). In the other, the system (called System 3) was similar to System 2, but 40 (out of the 100) PMMA chains were replaced by Py-PMMA-Py chains ($DP_n$ 30). Both simulations were executed at a relatively high temperature (melt state, $T=550K$ and $P=1atm$) in order to accelerate dynamics. GS agglomeration was observed in System 2 (see Figure S6), but not in System 3. Indeed, even after considerable simulation time (approximately 130 ns), all GS remained well separated and dispersed in the matrix. This is demonstrated in Figure 4a,b showing the time evolution of the distance between the centers-of-mass of all GS pairs present in the simulation cell. For System 3 (where a fraction of PMMA chains was substituted by Py-PMMA-Py chains), it is obvious that throughout the simulation, GS remain far apart with no tendency to approach each other and form π-π stacks. In contrast, in System 2 (no Py-PMMA-Py chains present), the distance of the centers-of-mass of GS pairs 1-2 and 2-3 is seen to reach very quickly (after approximately 3 ns) the value of 3.4 Å (i.e., the equilibrium distance of successive GS layers in a graphite flake), indicating graphene self-assembly in the matrix.

**Figure 4.** Time evolution of the distance between the centers-of-mass of all pairs of GS in the course of the MD simulations with Systems 2 (a) and 3 (b). (c) Time evolution of the % percentage of the various Py-PMMA-Py chain conformations (dangling ends, loops, bridges, free chains) in the course of the simulation. (d) Relative population of the various conformations in Systems 1 (presence of solvent) and 3 (presence of PMMA chains).

The various characteristic Py-PMMA-Py chain conformations (dangling ends, loops, bridges and free chains) appeared again in System 3 (see Figure S7), despite the presence of a rather large percentage of non-functionalized PMMA chains which hinders pyrene molecules from exploring their surroundings and adsorbing on the available GS surfaces. Figure 4c demonstrates the time evolution of these conformations that resemble quite well those obtained in solution (System 1). It seems that, a rather large number of adsorbed structures forms again in System 3. Figure 4d depicts the average population at steady state of each of the four different types of Py-PMMA-Py conformations for System 3 that, however, differ from those shown in Figure 1g for System 1 due to the very slow chain dynamics in System 3 (a rather dense system, since no solvent is present). The most probable chain conformation (amounting to ~50%) is that of dangling ends. It seems that the rest of the pyrene-end groups wander around for a rather considerable amount of time before coming across a GS surface to adsorb to. We also see that approximately 25% of Py-PMMA-Py chains form bridges, ~10% form loops, and the rest (~15%) are Py-PMMA-Py free chains.

Finally, encouraged by the simulation results, we attempted to produce G/Py-PMMA-Py dispersions in the ionic liquid 1-Ethyl-3-methylimidazolium bis(trifluoro methylsulfonyl) imide, [Emim][Tfsi] through several methods (see SI, Figure S9). For instance, G/Py-PMMA-Py hybrids, stably dispersed in $CHCl_3$, were successfully transferred to [Emim][Tfsi] by evaporating the solvent and replaced by the IL, followed by bath sonication (several hours) and then tip sonication at 40% for 60 min. The graphene concentration was determined at 0.3 mg/ml after centrifugation at 2000 rpm for 30 min. This preliminary result opens the possibility to fabricate ion gels by using the G/Py-PMMA-Py hybrids as the network forming material.

In conclusion, we have demonstrated a new methodology for the effective non-destructive liquid exfoliation of graphene toward nanocomposite fabrication. The designed pyrene-capped PMMA used as the dispersing agent led to stable graphene dispersions with relatively high concentrations of few-layered graphene sheets in chloroform. The suggested telechelic topology proved considerably a more efficient dispersing agent than the mono-functional Py-PMMA and the non-functionalized PMMA. More importantly, the atomistic simulations showed excellent agreement with the experimental results and revealed the type (and their time evolution) of the various GS/polymer conformations. Interestingly, the simulation predicts no agglomeration of GS non-covalently functionalized by Py-PMMA-Py in the PMMA matrix (composite). The identified conformations between the GS and the polymer (revealed either in solution or in the composite melt) are expected to have a profound effect upon the mechanical performance of solid or soft (e.g., ion gels) nanocomposites. It seems that the combination of the experiment with detailed atomistic simulations constitutes a powerful tool for designing graphene/functional polymer hybrids for optimizing mechanical reinforcement and opens the possibility for the production of graphene-based nanocomposites with tailor-made mechanical properties. Further work is in progress both in simulation (to predict optimum mass ratio of GS/Py-PMMA-Py/PMMA) and fabrication of graphene/PMMA reinforced composites as well as graphene/PMMA/ionic liquid gels (useful for electrochemical devices). Moreover, we plan to address issues related with the impact that the dispersing agents have on the elastic modulus of the nanocomposites, the degree of flexibility of graphene sheets and of the dispersing agents on these properties, and how this affects the type and strength of network conformations identified here. Additional issues to consider refer to the effect of the size of graphene sheets on their interaction(s) with the dispersing agents and the relative binding affinity of the dispersing agent onto graphene in the loop, extended loop, bridge, and dangling conformations using lower-level (quantum mechanical) simulations.

## ASSOCIATED CONTENT

**Supporting Information**.
Synthetic protocols, characterization of the polymers, MD simulation and exfoliation process (PDF).

## AUTHOR INFORMATION

**Corresponding Author**
*E-mail: ct@chemeng.upatras.gr
Fax: +30 2610 997266, Tel: +30 2610 969531## ACKNOWLEDGMENT

This work was supported by the European Union (European Social Fund-ESF) and Greek national funds through the research Funding Program: ERC-10 "Deformation, Yield and Failure of Graphene and Graphene-based Nanocomposites". The financial support of the Graphene FET Flagship (Grant Agreement No: 604391) is also acknowledged by CG. ENS and VGM acknowledge support through the project 'Multiscale Simulations of Complex Polymer Systems' (MuSiComPS) by the Limmat Foundation, Zurich, Switzerland. The computational results of this research were achieved using the PRACE-3IP project (FP7 RI-312763) resource SZEGED based in Hungary at NIIF. ENS and VGM feel indebted to Mr. Ioannis Liabotis from GR-NET, Greece, for his support on several technical aspects of the work.

## REFERENCES

1   Geim, A.K.; Novoselov, K.S. *Nat. Mater*. **2007**, 6, 183-91.
2   Ramanathan, T.; Abdala, A.A.; Stankovich, S.; Dikin, D.A.; Herrera-Alonso, M.; Piner R.D.; Adamson, D.H.; Schniepp H.C.; Chen, X.; Ruoff, R.S.; Nguyen, S.T.; Aksay, I.A.; Prud'homme, R.K.; Brinson, L.C. *Nat. Nanotechnol*. **2008**, 3, 327-31.
3   Li, X.G.; McKenna, G.B. *Acs Macro Lett*. **2012**, 1, 388-91.
4   Coleman, J.N. *Acc. Chem. Res*. **2013**, 46, 14–22.
5   Hernandez, Y.; Nicolosi, V.; Lotya, M.; Blighe, F.M.; Sun, Z.; De, S.; McGovern, I.T.; Holland, B.; Byrne, M.; Gun'Ko, Y.K.; Boland, J.J.; Niraj, P.; Duesberg, G.; Krishnamurthy, S.; Goodhue, R.; Hutchison, J.; Scardaci, V.; Ferrari, A.C.; Coleman, J.N. *Nat. Nanotechnol*. **2008**, 3, 563-68.
6   Kim, H.; Macosko, C.W. *Macromolecules* **2008**, 41, 3317-27.
7   Kim, H.; Miura, Y.; Macosko, C.W. *Chem. Mater.* **2010**, 22, 3441-50.
8   Layek, R.K.; Nandi, A.K. *Polymer* **2013**, 54, 5087-5266.
9   Parviz, D.; Das, S.; Ahmed, H.S.T.; Irin, F.; Bhattacharia, S.; Green, M.J. *ACS Nano* **2012**, 6, 8857-67.
10  Liu, J.; Tang, J.; Gooding, J.J. J. *Mater. Chem*. **2012**, 22, 12435-52.
11  Zheng, X.; Xu, Q.; Li, J.; Li, L.; Wei, J. *RSC Adv*. **2012**, 2, 10632-38.
12  Li, L.; Zheng, X.; Wang, J.; Sun, Q.; Xu, Q. *ACS Sustainable Chem. Eng.*, **2013**, *1*, 144–151.
13  Paek, K.; Yang, H.; Lee, J.; Park, J.; Kim, B.J. *ACS Nano*, **2014**, 8, 2848-56.
14  Xu, Y.; Bai, H.; Lu, G.; Li, C.; Shi, G.Q. *J. Am. Chem. Soc*. **2008**, 130, 5856-7.
15  Lee, D. W.; Kim, T.; Lee, M. *Chem. Commun*. **2011**, 47, 8259-61.
16  Mann, J. A.; Rodríguez-López, J.; Abruna, H. D.; Dichtel, W. R. *J. Am. Chem. Soc*. **2011**, 133, 17614-17.
17  Hu, K.; Kulkarni, D.D.; Choi, I.; Tsukruk, V.V. *Prog. Polym. Sci*. **2014**, 39, 1934-72.
18  Skountzos, E.N.; Anastassiou, A.; Mavrantzas, V.G.; Theodorou, D.N. *Macromolecules*, **2014**, 47, 8072-88.
19  Plimpton, S.J. *J. Comp. Phys.* **1995**, 117, 1-19. See also: http://www.sandia.gov/~sjplimp/lammps.html.
20  Xu, L.Y.; Yang, X.N. *J. Colloid Interf. Sci*. **2014**, 418, 66-71.
21  Zhang, Y.; Liu, C.; Shi, W.; Wang, Z.; Dai, L.; Zhang, X. *Langmuir* **2007**, 23, 7911-5.
22  Imaizumi, S.; Kokubo, H.; Watanabe, M. *Macromolecules* **2012**, 45, 401-409.
23  Paton, K.R.;   Varrla, E.; Backes, C.; Smith, R.J.; Khan, U.; O'Neil, A.; Boland, C.; Lotya, M.; Istrate, O.M.; King, P.; Higgins, T.; Barwich, S.; May, P.; Puczkarski, P.; Ahmed, I.; Moebius, M.; Petterson, H.; Long, E.;

Coelho, J.; O'Brien, S.E.; McGuire, E.K.; Sanchez, B.M.; Duesberg, G.S.; McEvoy, N.; Pennycook, T.J.; Downing, C.; Crossley, A.; Nicolosi V.; Coleman, N. *J. Nat. Mater.* **2014**, 13, 624-30.
24 Ciesielski, A.; Samorì, P. *Chem. Soc. Rev.* **2014**, 43, 381-98.
25 Wang, H.; Chen, Z.; Xin, L.; Cui, J.; Zhao S.; Yan, Y. *J. Polym. Sci., Part A: Polym. Chem.* **2015**, DOI: 10.1002/pola.27675.
26 Kim, H.; Kobayashi, S.; AbdurRahim, M.A.; Zhang, M.L.J.; Khusainova, A.; Hillmyer, M.A.; Abdala, A.A.; Macosko, C.W. *Polymer* **2011**, 52, 1837-46.
27 Barroso-Bujans, F.; Cerveny, S.; Verdejo, R.; Val,J.J.; Alberdi, J.M.; Alegra A.; Colmenero, J. *Carbon* **2009**, 48, 1079-87.
28 Fasolino, A.; Los, J. H.; Katsnelson, M.I. *Nat. Mater.* **2007**, 6, 858-61.
29 Ferrari, A.C.; Meyer, J.C.; Scardaci, V.; Casiraghi, C.; Lazzeri, M.; Mauri, F.; Piscanec, S.; Jiang, D.; Novoselov, K.S.; Roth, S.; Geim, A.K. *Phys. Rev. Lett.* **2006**, 97, 187401.
30 Malarda, L. M.; Pimentaa, M. A.; Dresselhausb, G.; Dresselhausc, M.S. *Physics Reports* **2009**, 473, 51-88.
31 Gong, L.; Young, R.J.; Kinloch, I.A.; Riaz, I.; Jalil, R.; Novoselov, K.S. *ACS Nano* **2012**, 6, 2086-95.